# A New Direction for First-Principles Device Simulations

Yong-Hoon Kim* and Ryong-Gyu Lee

*School of Electrical Engineering, Korea Advanced Institute of Science and Technology (KAIST), 291 Daehak-ro, Yuseong-gu, Daejeon 34141, Korea.*

E-mail: y.h.kim@kaist.ac.kr

**Abstract**

The continued miniaturization of semiconductor devices, represented by Moore's law, has reached the atomic scale limit, requiring nanoscale quantum mechanical effects to be included in device simulations without empirical parameters. For this purpose, a method that combines density functional theory (DFT) and non-equilibrium Green's function (NEGF) theory has been established as the standard approach for atomic-scale device simulations. However, the DFT-NEGF scheme has several inherent weaknesses due to the underlying Landauer or grand canonical viewpoint. To overcome these challenges, over the years, we have developed an alternative approach, the multispace constrained-search DFT (MS-DFT) formalism. The central starting point of this development is the replacement of the Landauer picture by the multispace excitation viewpoint or the mapping of the quantum transport process to the energy- and position-space electron excitation within a (micro)canonical ensemble device model. This change in the viewpoint leads to several fundamental advantages of MS-DFT over its DFT-NEGF counterparts, such as the variational determination of non-equilibrium total energy, the explicit extraction of quasi-Fermi level distributions, and the possibility of faithfully modeling finite 2D electrodes. In this article, we highlight the key features and applications of the MS-DFT formalism.

## 1. INTRODUCTION

Over the past half-century, the semiconductor industry has continuously extended Moore's law and miniaturized metal-oxide-semiconductor field effect transistors (MOSFETs) through many scientific breakthroughs and engineering innovations. Behind this success lie process and device technology computer-aided design (TCAD) simulations, which provided a deeper understanding of materials and device characteristics and allowed the design of complex circuits. Currently, the shrinkage of device dimensions to nanometer-level technology nodes and beyond demands that the evolution of TCAD simulations be able to reliably describe various atomic-scale quantum effects emerging in extremely miniaturized semiconductor devices.

In the last century, density functional theory (DFT) was established as the standard approach for first-principles electronic structure calculations for materials in equilibrium. Accordingly, for the past two decades, the method that combines DFT and the non-equilibrium Green's function (NEGF) formalisms has been established as the standard approach for non-equilibrium electronic structure and quantum transport calculations [1, 2]. The DFT-NEGF approach involves the Landauer picture, which views the device as an open quantum system in which a channel is coupled to semi-infinite left and right electrodes [3-5]. The introduction of this Landauer picture, however, imposes several limitations on the DFT-NEGF. For example, since the Landauer picture is based on the grand canonical ensemble description, DFT-NEGF requires electrodes to be repeated semi-infinitely along the transport direction and cannot handle finite-sized electrodes such as graphene. More importantly, the variational total energy, which represents the central quantity and the source of the power of the DFT [6], is unavailable within the DFT-NEGF [7].

In the past decade or so, as an alternative to the DFT-NEGF approach, we developed the multi-space constrained-search DFT (MS-DFT) formalism [8-10]. The starting point of this development was, as implied above, replacing the Landauer picture for quantum transport by the multi-space excitation viewpoint (Fig. 1(a)). Specifically, we adopt the (micro)canonical ensemble or finite-electrode models and map the finite-bias quantum transport process to the (borrowing the MOSFET terminology) drain-to-source electron excitation counterpart. In this article, after briefly sketching the foundations of the MS-DFT formalism, we will describe several examples of its applications, including the variational determination of non-equilibrium total energy [9], the definition of non-equilibrium adsorption energy [11], the extraction of quasi-Fermi level splitting information [9], the faithful modeling of finite-electrode systems [12, 13], and the consistent treatment of quantum transport and optical excitation processes.

## 2. FOUNDATIONS OF MS-DFT

### 2.1 DFT-NEGF

As a reference, we first briefly discuss the NEGF formalism based on the Landauer picture. The Landauer viewpoint [3, 5] relies on several physical assumptions that include the presence of a geometrically or energetically narrow nanoconstriction for the channel (*C*) sandwiched between the left reservoir (*L*) and right reservoir (*R*) (Fig. 1(b)). Under the non-equilibrium condition established via a finite applied bias voltage $V_b = (\mu_L - \mu_R)/e$, where $\mu_{L/R}$ represent the *L/R* electrochemical potentials, the *L* and *R* reservoirs remain in local equilibrium and are characterized by Fermi-Dirac functions:

$$f^{L/R}(E) \equiv f(E - \mu_{L/R}) = \frac{1}{1 + \exp[(E - \mu_{L/R})k_B T}  \quad (1)$$

We emphasize that this can be considered a condition that guarantees accurate DFT-NEGF calculations. As an example, the $L$ and $R$ "reservoirs" typically do not correspond to physical metal "electrodes" because the interfacial "lead" regions of electrodes partly accommodate voltage drops and should be considered parts of $C$.

The NEGF method is formulated in terms of two Green's functions that serve two distinct purposes [2, 14]. First, the open boundary conditions are addressed with the retarded Green's functions

$$\mathbf{G}(E) = [E\mathbf{S} - \mathbf{H} - \mathbf{\Sigma}_L(E) - \mathbf{\Sigma}_R(E)]^{-1}, \quad (2)$$

where $\mathbf{\Sigma}_{L(R)}$ is the self-energy matrix and describes the coupling between $C$ and the semi-infinite $L/R$ reservoirs. Second, the non-equilibrium electronic structure or density matrix is evaluated in terms of the lesser Green's function $\mathbf{G}^<$ or electron correlation function $\mathbf{G}^n$ according to

$$\mathbf{G}^n(E) \equiv -i\mathbf{G}^<(E) \approx \mathbf{G}(E)\mathbf{\Gamma}_L(E)\mathbf{G}^\dagger(E)f_L(E)$$
$$+ \mathbf{G}(E)\mathbf{\Gamma}_R(E)\mathbf{G}^\dagger(E)f_R(E), \quad (3)$$

where $\mathbf{\Gamma}_{L(R)}$ is the $L(R)$ electrode-induced broadening matrix defined as the anti-Hermitian part of the self-energy matrix $\mathbf{\Gamma}_{L(R)} = i(\mathbf{\Sigma}_{L(R)} - \mathbf{\Sigma}^\dagger_{L(R)})$.

## 2.2 MS-DFT

Given the fundamental difference between the Landauer and multi-space excitation viewpoints for quantum transport, the MS-DFT formalism is characterized in terms of several unique features and established through the following steps (Fig. 1(c)):

*Step 1. Microcanonical or canonical ensemble.* In contrast to the grand canonical ensemble associated with the Landauer picture and DFT-NEGF, the multi-space excitation viewpoint and MS-DFT employ the microcanonical (at zero temperature) or canonical (at finite temperature) ensemble and replace semi-infinite electrodes with finite metal slabs.

*Step 2. Partitioning of wavefunctions.* The most notable common feature between DFT-NEGF and MS-DFT is likely the division of the junction system into the $L$, $C$, and $R$ regions. However, rather than the Green's functions $\mathbf{G}$ and $\mathbf{G}^n$, the wavefunctions $\Psi$ are the main ingredients of MS-DFT, and the spatial origins of $\Psi$ are traced to $L$ or $C$ or $R$. At the zero-bias equilibrium limit, $\Psi$ collectively provides the ground-state density $\rho_0(\vec{r}) = \rho_0^L(\vec{r}) + \rho_0^C(\vec{r}) + \rho_0^R(\vec{r})$.

*Step 3. Multi-space constrained search.* We now view the quantum *transport* process driven by a finite applied bias voltage $V_b = (\mu_L - \mu_R)/e$ as the multi-space *excitation* process, which is composed of (1) the position-space jump from the drain reservoir $R$ to the source reservoir $L$ and (2) the energy-space jump from $\mu_R$ to $\mu_L$. A constrained-search procedure is then used to minimize the total energy with respect to the multi-space excited states with density $\rho_k$. Specifically, given the ground state with total energy $E_0$ and density $\rho_0$, the governing equation of MS-DFT becomes a constrained search of the total energy minimum of the spatially excited state $k$ with density $\rho_k(\vec{r}) = \rho_k^L(\vec{r}) + \rho_k^C(\vec{r}) + \rho_k^R(\vec{r})$:

$$E_k = \min_\rho \{\int v(\vec{r})\rho(\vec{r})d^3\vec{r} + F[\rho_k^L, \rho_k^C, \rho_k^R, \rho_0]\}$$
$$= \int v(\vec{r})\rho(\vec{r})d^3\vec{r} + F[\rho_k, \rho_0], \quad (4)$$

with the universal functional

$$F[\rho_k, \rho_0] = \min_{\Psi^{L/C/R} \to \rho_k} \langle \Psi^{L/C/R} | \hat{T} + \hat{V}_{ee} | \Psi^{L/C/R} \rangle, \quad (5)$$

where the spatially resolved $\Psi^{L/C/R}$ satisfy the bias constraint of $eV_b = (\mu_L - \mu_R)$ and are orthogonal to the first $k$-1 excited states.

### 2.2.1 Kohn-Sham electronic structure calculations.

For practical non-equilibrium electronic structure calculations, as in the ground-state DFT formalism, we then employ the single-electron Kohn-Sham (KS) mapping and solve the following KS equations:

$$[\hat{h}^0_{KS} + \Delta v_{Hxc}(\vec{r})]\psi_i(\vec{r}) = \varepsilon_i \psi_i(\vec{r}), \quad (6)$$

to obtain $\rho_k$ and $E_k$ within the partitioning of the $L/C/R$ regions and the constraint of $eV_b = (\mu_L - \mu_R)$. Here, $\hat{h}^0_{KS}$, $\Delta v_{Hxc}(\vec{r})$, $\psi_i(\vec{r})$, and $\varepsilon_i$ represent the ground-state KS Hamiltonian, bias-induced modification of the Hamiltonian, KS eigenstates, and KS eigenvalues, respectively. Here, the rule for the occupation $f_i^C$ of channel KS states $\psi_i^C$ needs to be established. This corresponds to the identification of quasi-Fermi levels (QFLs), and once $f_i^C$ is available one can compute the $C$ density

$$\rho_i^C(\vec{r}) = \sum_i f_i^C |\psi_i^C(\vec{r})|^2. \quad (7)$$

Owing to the practical insignificance of the exchange-correlation potential term $\Delta v_{xc}$ to the build-up of the voltage drop, typically, only the classical Hartree electrostatic term needs to be considered in the analysis. It can also be noted that $\Delta v_H$ is directly related to the Landauer residual-resistivity dipole

$$\Delta\rho = \rho(V_b) - \rho(0).$$

*2.2.2 Quantum transport calculations.*

We emphasize that, while the non-equilibrium electronic structure of a device (which includes *L* and *R* as well as *C*) is computed self-consistently by solving the non-equilibrium KS equations, $\Sigma_{L(R)}$ or the information from separate bulk crystal calculations was not introduced. Instead, only after the nonequilibrium electronic structure is variationally determined, the quantum transport properties are derived as a "postprocessing" step by recovering the Landauer picture. Then, by constructing a self-energy matrix $\Sigma_{L(R)}$ and retarded Green's functions **G**, the transmission function can be calculated based on the matrix Green's function formalism (MGF) [15, 16]:

$$T(E;V_b) = Tr[\Sigma_{L(R)}\Gamma_L(E;V_b)G(E;V_b)\Gamma_R(E;V_b)G^\dagger(E;V_b)]. \quad (8)$$

The current-bias voltage ($I$–$V_b$) characteristic is finally calculated via the Landauer-Büttiker formula [2, 14],

$$I(V_b) = \frac{2e}{h}\int_{-\infty}^{+\infty}T(E;V_b)[f(E-\mu_R) - f(E-\mu_R)]dE. \quad (9)$$

As evident from the above, a key feature of MS-DFT is the separation of the variational determination of the non-equilibrium electronic structure and the post-processing calculation of the quantum transport properties. The electron correlation function $\mathbf{G}^n$ is not introduced at all within MS-DFT calculations.

## 3. APPLICATION EXAMPLES

By utilizing several application examples, we now highlight several key features of MS-DFT.

### 3.1 Non-equilibrium total energy

A central discriminating feature between the DFT-NEGF and MS-DFT calculations is the availability of the non-equilibrium total energy. Beyond the fundamental theoretical level, this also has enormous practical implications in that the variational determination of the total energy is the key to guaranteeing reliable self-consistent DFT and related ground-state electronic structure calculations. In Fig. 2, for a molecular junction in which a benzenedithiolate (BDT) molecule is stretched between Au(111) electrodes, we show the convergence behavior of the $V_b = 0.8V$ non-equilibrium total energy as the basis set size increases. Note that, in contrast to the total energy, the electrical current is not a variational quantity and exhibits an increasing trend with the basis set size and even small fluctuations beyond the level where the total energy has converged (DZP) (Fig. 1(c), red square). This result implies the practical advantages of MS-DFT calculations, which allow systematic checking of the convergence of non-equilibrium self-consistent electronic structure calculations.

### 3.2 QFL distributions

The concept of the quasi-Fermi level (also referred to as the electrochemical potential or as the *imref*, which is "fermi" spelled backward) and the splitting of electron and hole QFLs represents a cornerstone of semiconductor device physics and typically appears even in undergraduate-level textbooks. However, atomic-scale characterization of QFLs has remained a major challenge even 70 years after their inception by William Shockley [17]. As explained above, the explicit calculations of $\psi_i^C$ and $f_i^C$ or QFLs are part of the MS-DFT calculation algorithm. In Fig. 3, we show the calculated QFLs of two molecular junctions constructed by sandwiching hexandithiolate (HDT) and hexatrienedithiolate (HTDT) molecules between the Au(100) electrodes. For the insulating HDT junction, the QFLs in the channel region are missing within the bias window (Fig. 3(a)). In this case, the electrostatic potential ($\Delta v_H = v_H(V_b) - v_H(0)$) decreases linearly, and the corresponding Landauer residual-resistivity dipole ($\Delta\rho = \rho(V_b) - \rho(0)$) is uniformly distributed across the molecule. For the semiconducting HTDT case, however, the QFLs from the left and right electrodes penetrate the channel region (Fig. 3(b)). This leads to QFL splitting, which is accompanied by nonlinear $\Delta v_H$ and asymmetric $\Delta\rho$ formations.

### 3.3 Finite-sized low-dimensional electrodes

Atomically thin 2D van der Waals (vdW) materials are promising building blocks for next-generation electronic devices. In particular, significant effort has been devoted to identifying device architectures that maximally leverage the unique quantum properties emerging from diverse combinations of 2D vdW materials. Obviously, this is an area where first-principles process and device TCAD simulations can play a crucial role, and the DFT-NEGF has been extensively utilized for this purpose. However, as schematically shown in Figs. 4(a) and 4(b), the DFT-NEGF has intrinsic limitations in simulating, e.g., 2D vdW heterojunctions based on graphene electrodes. The Landauer framework behind the DFT-NEGF requires electrodes to be repeated semi-infinitely along the electron transport direction, forcing one to replace graphene with graphite or to adopt nanoribbon channel models. Accordingly, such 2D vertical vdW FETs could previously be studied only by semiclassical approaches such as the Bardeen transfer Hamiltonian formalism [18, 19].

On the other hand, MS-DFT, which adopts a microcanonical viewpoint, does not suffer from this issue. MS-DFT simulations for Au/vacuum/graphene/few-layer hexagonal boron nitride (hBN)/graphene heterostructure models show that the experimentally observed symmetric negative differential resistance (NDR) behavior [19] originates from atomic defects and that the details of these defects can significantly affect device characteristics (Fig. 4(d)).

The possibility of modeling finite-sized electrodes allows the incorporation of additional gate electrode(s) within the vertical FET model. In collaboration with an experimental team, we studied the emergence of negative differential transconductance (NDT) from a p-i-n WSe$_2$ double lateral homojunction transistor, in which n- and p-type doping were induced by polymethyl methacrylate and Au$_2$Cl$_6$, respectively [13]. MS-DFT simulations revealed that the exceptional NDT performance results from the modulation of trap-assisted tunneling and nonlinear 2D band shifts in response to gate control.

## 4. SUMMARY AND OUTLOOK

In this article, we sketched the foundations of the MS-DFT formalism and highlighted its key features via several application examples. Rather than simply representing an alternative to DFT-NEGF, we explained that MS-DFT, in fact, goes beyond DFT-NEGF on several fronts. It was emphasized that this was possible by replacing the underlying Landauer framework for quantum transport by the multi-excitation viewpoint. We close this article by providing several promising follow-up research directions.

### 4.1 Electric enthalpy and non-equilibrium atomic forces

We believe that the variational non-equilibrium total energy highlighted above and the associated atomic forces represent key components of MS-DFT that will play a crucial role in future process and device TCAD simulations. For this purpose, as an initial step, we formulated a theory of electric enthalpy, which is the appropriate thermodynamic potential for electrified interfaces [11]. As the next step, for interfaces relevant for various device applications, we are carrying out non-equilibrium migration barrier calculations and molecular dynamics simulations.

### 4.2 Optoelectronic device simulations

Within the scope of the multi-excitation viewpoint or quantum transport-to-optical excitation mapping, the variational time-independent excited-state DFT or □SCF formalism might be viewed as the energy-space-only excitation limit of MS-DFT [20, 21]. Accordingly, we expect that MS-DFT will provide a natural foundation for describing concurrent transport and optical excitation processes in optoelectronic devices, which include electron-phonon and electron-photon couplings in a coherent manner [22].

### 4.3 Advanced exchange-correlation functionals

Due to the self-interaction errors in the exchange-correlation functionals, MS-DFT and DFT-NEGF calculations based on the typical local density approximation and generalized gradient approximation result in large quantitative errors in the calculated current values [23]. Another promising extension of MS-DFT we are currently exploring is the incorporation of orbital-based exchange-correlation functionals [24-26]. While this is a non-trivial task once the NEGF route is adopted, the utilization of orbital-based exchange-correlation functionals is straightforward within MS-DFT and is expected to provide an accurate yet efficient solution to this challenge.

### 4.4 Multi-scale device simulations

Finally, in view of simulating large device models composed of tens of thousands of atoms or more, we are currently exploring approaches to carry out MS-DFT calculations for atomistic device models electrostatically embedded within a polarizable continuum dielectric environment described at the effective mass approximation level [25, 27, 28]. We hope to update these exciting developments in the near future.

## ACKNOWLEDGEMENTS

This work was supported by the National Research Foundation of Korea (2022K1A3A1A91094293, RS-2023-00253716).

## REFERENCES

[1] A. Pecchia and A. D. Carlo, "Atomistic theory of transport in organic and inorganic nanostructures," *Rep. Prog. Phys.,* vol. 67, no. 8, pp. 1497-1561, 2004, doi: 10.1088/0034-4885/67/8/r04.

[2] M. Di Ventra, *Electrical Transport in Nanoscale Systems*. Cambridge, UK: Cambridge University Press, 2008.

[3] R. Landauer, "Electrical transport in open and closed systems," *Z. Phys. B: Condens. Matter,* vol. 68, no. 2-3, pp. 217-228, 1987, doi: 10.1007/bf01304229.

[4] R. Landauer, "Conductance determined by transmission: probes and quantised constriction resistance," *J. Phys. Condens. Matter,* vol. 1, no. 43, pp. 8099-8110, 1989, doi: 10.1088/0953-8984/1/43/011.

[5] R. Landauer, "Conductance from transmission: common sense points," *Physica Scripta,* vol. T42, pp. 110-114, 1992, doi: 10.1088/0031-8949/1992/t42/020.

[6] W. Kohn, "Nobel Lecture: Electronic structure of matter—wave functions and density functionals," *Rev. Mod. Phys.,* vol. 71, no. 5, pp. 1253-1266, 1999, doi: 10.1103/RevModPhys.71.1253.

[7] Z. Yang, A. Tackett, and M. Di Ventra, "Variational and nonvariational principles in quantum transport calculations," *Phys. Rev. B,* vol. 66, no. 4, 2002, doi: 10.1103/PhysRevB.66.041405.

[8] H. S. Kim and Y.-H. Kim, "Constrained-search density functional study of quantum transport in two-dimensional vertical heterostructures," *arXiv preprint,* pp. arXiv:1808.03608 [cond-mat.mes-hall], 2018, doi: arxiv.org/abs/1808.03608.

[9] J. Lee, H. S. Kim, and Y.-H. Kim, "Multi-space excitation as an alternative to the Landauer picture for non-equilibrium quantum transport," *Adv. Sci.,* vol. 7, no. 16, p. 2001038, Aug 2020, doi: 10.1002/advs.202001038.

[10] J. Lee, H. Yeo, and Y.-H. Kim, "Quasi-Fermi level splitting in nanoscale junctions from ab initio," *Proc. Natl. Acad. Sci. U. S. A.,* vol. 117, no. 19, pp. 10142-10148, May 12 2020, doi: 10.1073/pnas.1921273117.

[11] J. Lee, H. Yeo, R.-G. Lee, and Y.-H. Kim, "Ab initio theory of the nonequilibrium adsorption energy," *npj Comput. Mater.,* vol. 10, no. 1, 2024, doi: 10.1038/s41524-024-01242-5.

[12] T. H. Kim, J. Lee, R.-G. Lee, and Y.-H. Kim, "Gate- versus defect-induced voltage drop and negative differential resistance in vertical graphene heterostructures," *npj Comput. Mater.,* vol. 8, no. 1, 2022, doi: 10.1038/s41524-022-00731-9.

[13] H. Son *et al.*, "Emergence of multiple negative differential transconductance from a WSe2 double lateral homojunction platform," *Appl. Surf. Sci.,* vol. 581, 2022, doi: 10.1016/j.apsusc.2021.152396.

[14] S. Datta, *Electronic Transport in Mesoscopic Systems*. World Scientific 1995.

[15] Y.-H. Kim, S. S. Jang, Y. H. Jang, and W. A. Goddard, 3rd, "First-principles study of the switching mechanism of [2]catenane molecular electronic devices," *Phys. Rev. Lett.,* vol. 94, no. 15, p. 156801, Apr 22 2005, doi: 10.1103/PhysRevLett.94.156801.

[16] Y.-H. Kim, J. Tahir-Kheli, P. A. Schultz, and W. A. Goddard, "First-principles approach to the charge-transport characteristics of monolayer molecular-electronics devices: Application to hexanedithiolate devices," *Phys. Rev. B,* vol. 73, no. 23, p. 235419, 2006, doi: 10.1103/PhysRevB.73.235419.

[17] W. Shockley, "The theory of p-n junctions in semiconductors and p-n junction transistors," *Bell System Technical Journal,* vol. 28, no. 3, pp. 435-489, 1949, doi: 10.1002/j.1538-7305.1949.tb03645.x.

[18] J. Bardeen, "Tunnelling from a many-particle point of view," *Phys. Rev. Lett.,* vol. 6, no. 2, pp. 57-59, 1961, doi: 10.1103/PhysRevLett.6.57.

[19] L. Britnell *et al.*, "Resonant tunnelling and negative differential conductance in graphene transistors," *Nat. Commun.,* vol. 4, p. 1794, 2013, doi: 10.1038/ncomms2817.

[20] A. Görling, "Density-functional theory beyond the Hohenberg-Kohn theorem," *Phys. Rev. A,* vol. 59, pp. 3359-3374, 1999, doi: 10.1103/PhysRevA.59.3359.

[21] M. Levy and Á. Nagy, "Variational density-functional theory for an individual excited state," *Phys. Rev. Lett.,* vol. 83, no. 21, pp. 4361-4364, 1999, doi: 10.1103/PhysRevLett.83.4361.

[22] S. Seo *et al.*, "An optogenetics-inspired flexible van der Waals optoelectronic synapse and its application to a convolutional neural network," *Adv. Mater.,* vol. 33, no. 40, p. e2102980, Oct 2021, doi: 10.1002/adma.202102980.

[23] Y.-H. Kim, H. S. Kim, J. Lee, M. Tsutsui, and T. Kawai, "Stretching-induced conductance variations as fingerprints of contact configurations in single-molecule junctions," *J. Am. Chem. Soc.,* vol. 139, no. 24, pp. 8286-8294, Jun 21 2017, doi: 10.1021/jacs.7b03393.

[24] Y.-H. Kim, M. Städele, and R. M. Martin, "Density-functional study of small molecules within the Krieger-Li-Iafrate approximation," *Phys. Rev. A,* vol. 60, no. 5, pp. 3633-3640, 1999, doi: 10.1103/PhysRevA.60.3633.

[25] Y.-H. Kim, I.-H. Lee, S. Nagaraja, J.-P. Leburton, R. Q. Hood, and R. M. Martin, "Two-dimensional limit of exchange-correlation energy functional approximations," *Phys. Rev. B,* vol. 61, no. 8, pp. 5202-5211, 2000, doi: 10.1103/PhysRevB.61.5202.

[26] Y.-H. Kim and A. Görling, "Excitonic optical spectrum of semiconductors obtained by time-dependent density-functional theory with the exact-exchange kernel," *Phys. Rev. Lett.,* vol. 89, no. 9, 2002, doi: 10.1103/PhysRevLett.89.096402.

[27] I.-H. Lee, K.-H. Ahn, Y.-H. Kim, R. M. Martin, and J.-P. Leburton, "Capacitive energies of quantum dots with hydrogenic impurity," *Phys. Rev. B,* vol. 60, no. 19, pp. 13720-13726, 1999, doi: 10.1103/PhysRevB.60.13720.

[28] H. Yeo, J. S. Lee, M. E. Khan, H. S. Kim, D. Y. Jeon, and Y.-H. Kim, "First-principles-derived effective mass approximation for the improved description of quantum nanostructures," (in English), *J. Phys. Mater.,* vol. 3, no. 3, Jul 2020, doi: ARTN 03401210.1088/2515-7639/ab9b61.

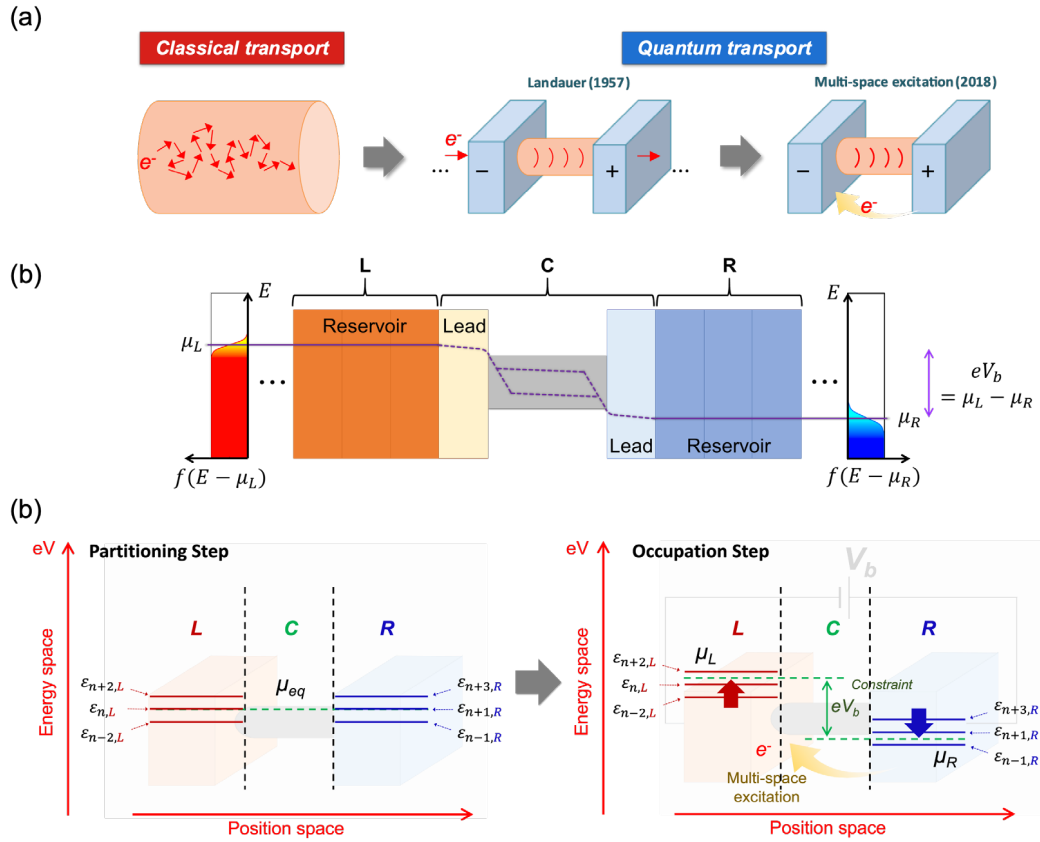

**Figure 1.** (a) Landauer vs. multi-space excitation viewpoints for quantum transport. (b) Schematic of the Landauer picture for a nanojunction. For the junction consisting of the channel (C), left (L) and right (R) reservoirs, the applied bias voltage $V_b$ is introduced as the boundary condition imposed on the L and R reservoirs remaining in equilibrium. (c) Schematic of the MS-DFT framework, which maps the L-to-R quantum transport process to R-to-L electron excitation. The computational procedure involves a partitioning step and an occupation step. Reprinted with permission from [9]. (b) and (c) © 2020 WILEY-VCH.

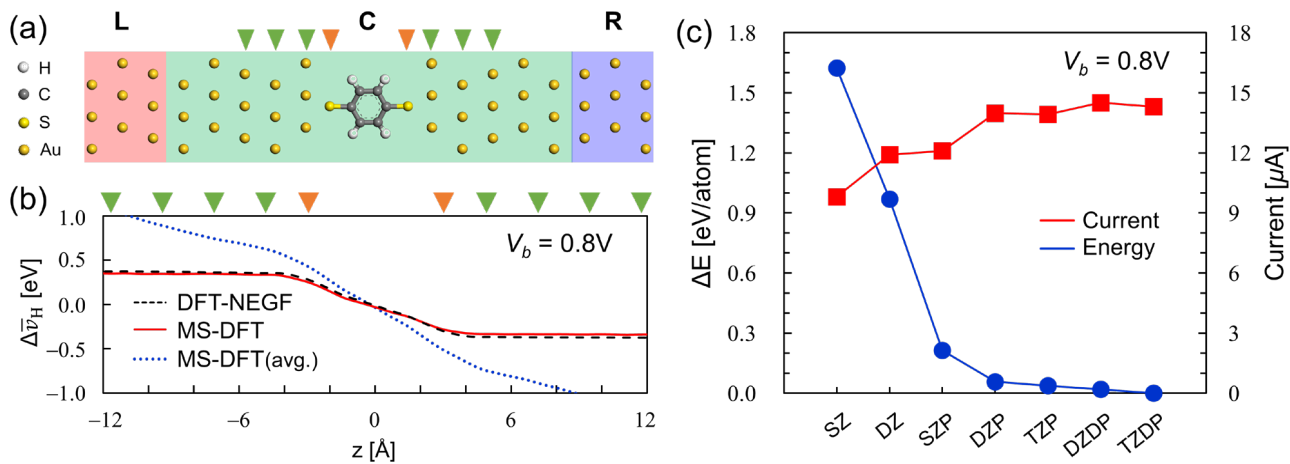

**Figure 2.** (a) BDT junction model. (b) Plane-averaged electrostatic potential drop $\Delta \bar{v}_H$ for $V_b$ = 0.8 V, calculated from the DFT-NEGF (dotted black line), MS-DFT following the occupation rule of (9) (red line), and MS-DFT following the averaged occupation rule: $(\mu_L + \mu_R)/2$ (dotted blue line), indicating that the DFT-NEGF and MS-DFT with the occupation rule of (9) are equivalent. (c) Convergence behavior of the nonequilibrium total energy (blue filled circles) and current (red filled squares) with respect to the basis-set level (SZ: single-$\zeta$; DZ: double-$\zeta$; SZP: single-$\zeta$ plus polarization; DZP: double-$\zeta$ plus polarization; TZP: triple-$\zeta$ plus polarization; DZDP: double-$\zeta$ plus double-polarization; TZDP: triple-$\zeta$ plus double-polarization). Reprinted with permission from [9]. © 2020 WILEY-VCH

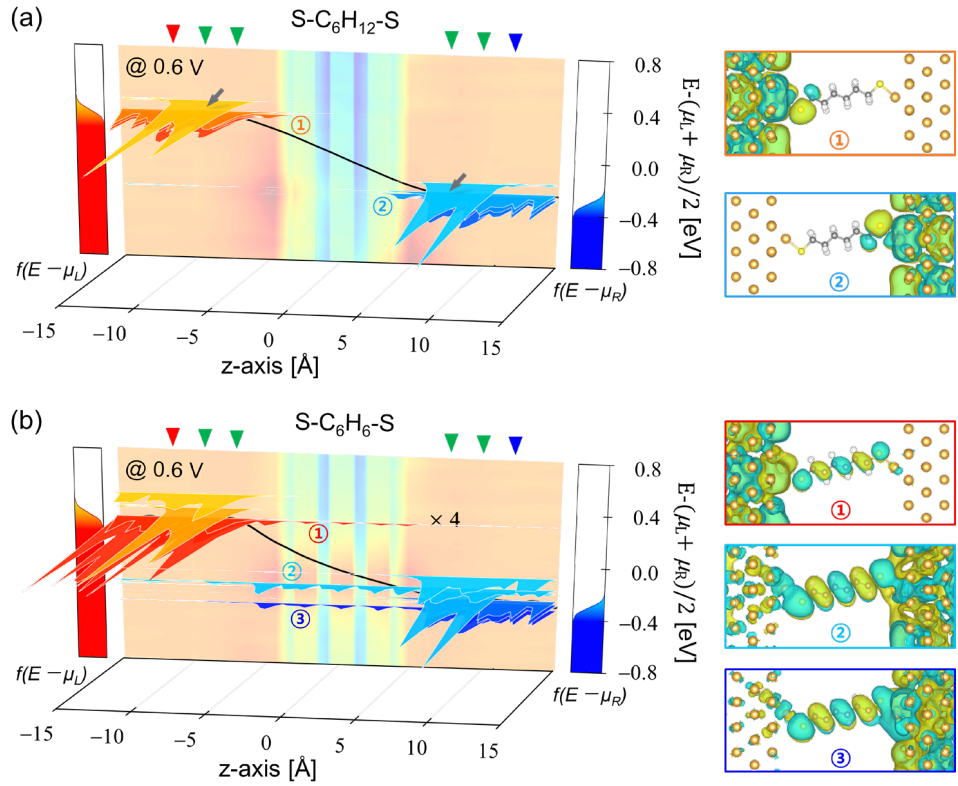

**Figure 3.** The non-equilibrium QFL profiles for the (a) HDT and (b) HTDT junctions at $V_b$ = 0.6 V obtained from the MS-DFT calculations, where the QFL wavefunctions are presented by color-coded distributions according to the electron occupation given by $f^L$ and $f^R$ for the L- (red) and R-originated (blue) states, respectively. The spatially resolved density of states is also shown as transparent background images in (a) and (b). The most prominent QFL wavefunctions within region C are also presented in the right panels (① and ② in (a) and ①, ②, and ③ in (b)). Reprinted with permission from [10]. © 2020 National Academy of Sciences.

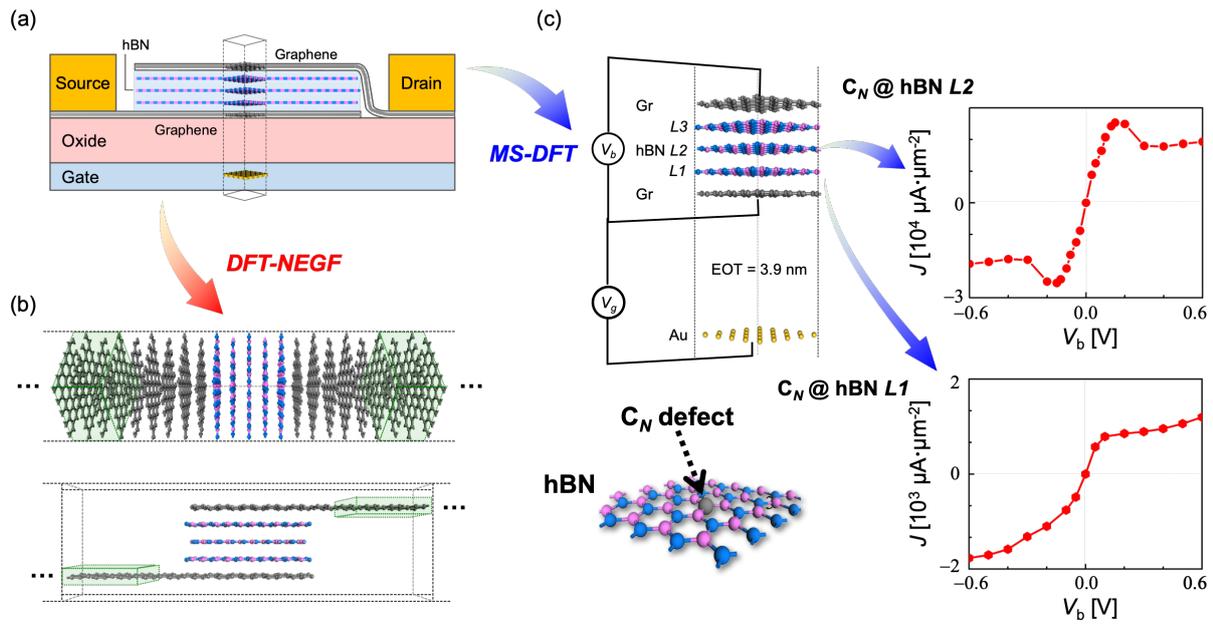

**Figure 4.** (a) Schematic of the graphene-based vertical FET device. (b) Within the DFT-NEGF, because it requires semi-infinite electrodes, one needs to either replace graphene with graphite or replace 2D hBN layers with hBN nanoribbon models. (c) On the other hand, MS-DFT can faithfully model graphene-based vertical FETs, including gate electrodes. (d) Simulation results showing the critical impact of atomic defects on device characteristics. Reprinted with permission from [12]. © 2022 Springer Nature.

9